\documentclass[twocolumn,showpacs,preprintnumbers,amsmath,amssymb]{revtex4}

\usepackage{graphicx}
\usepackage{dcolumn}
\usepackage{bm}
\usepackage[T1]{fontenc}

\begin{document}

\preprint{APS/123-QED}

\title{Noise properties of nanoscale YBCO Josephson junctions}

\author{D. Gustafsson}
\email{dagust@chalmers.se}
\author{F. Lombardi}
\author{T. Bauch}

\affiliation{Department of Microtechnology and Nanoscience, Chalmers
University of Technology, SE-412 96 Göteborg, Sweden}

\date{\today}

\begin{abstract}

We present electric noise measurements of nanoscale biepitaxial
YBa$_2$Cu$_3$O$_{7-\delta}$ (YBCO) Josephson junctions fabricated by
two different lithographic methods. The first (conventional)
technique defines the junctions directly by ion milling etching
through an amorphous carbon mask. The second (soft patterning)
method makes use of the phase competition between the
superconducting YBCO (Y123) and the insulating Y$_2$BaCuO$_5$ (Y211)
phase at the grain boundary interface on MgO (110) substrates. The
voltage noise properties of the two methods are compared in this
study. For all junctions (having a thickness of 100 nm and widths of
250-500 nm) we see a significant amount of individual charge traps.
We have extracted an approximate value for the effective area of the
charge traps from the noise data. From the noise measurements we
infer that the soft patterned junctions with a grain boundary (GB)
interface manifesting a large c-axis tunneling component have a
uniform barrier and a SIS like behavior. The noise properties of
soft patterned junctions having a GB interface dominated by
transport parallel to the ab-planes are in accordance with a
resonant tunneling barrier model. The conventionally patterned
junctions, instead, have suppressed superconducting transport
channels with an area much less than the nominal junction area.
These findings are important for the implementation of nanosized
Josephson junctions in quantum circuits.
\end{abstract}


\maketitle \textit{This document is the Accepted Manuscript version
of a Published Work that appeared in final form in Physical Review
B, copyright \copyright American Physical Society. To access the
final edited and published work see
\newline http://prb.aps.org/abstract/PRB/v84/i18/e184526}
\section{\label{sec:level1}Introduction}

Micrometer-sized grain boundary (GB) Josephson junctions (JJs) made
of High Critical Temperature Superconductors (HTS) are commonly used
for the realization of superconducting devices operated in a wide
temperature range up to the boiling temperature of liquid nitrogen.
A prominent example is the Superconducting QUantum Interference
Device (SQUID) for sensitive magnetic flux detection
\cite{Koelle1999}. Nevertheless, GB JJs are still a fundamental tool
for the exploration of the complex physics inherent to HTS materials
\cite{Alff1998,Tsuei2000,Jesper,Bauch2005,Bauch2006}. Their
implementation in superconducting circuits operated in the quantum
limit, such as quantum bits or single electron transistors, are
expected to give further useful hints on the unresolved nature of
superconductivity in HTS materials \cite{Bauch2009}. Recent advances
in the thin film technology and nano-fabrication of HTS made it
possible to observe macroscopic quantum phenomena in
YBa$_2$Cu$_3$O$_{7-\delta}$ (YBCO) biepitaxial grain boundary
Josephson junctions \cite{Bauch2005,Bauch2006} opening the way for
the realization of HTS quantum circuits. For typical JJ-based
devices, which operate in the quantum limit (typically at
temperatures below 100 mK), the requirements on junction critical
currents and capacitances are met for lateral dimensions on a length
scale of 100 nm \cite{Bauch2009,Gustafsson2010}.

The realization of reproducible HTS JJs at the nanoscale can also be
instrumental to fabricate sensors with a quantum limited sensitivity
like nano-SQUIDs, which can allow the detection of magnetic
nano-particles in a much wider temperature and magnetic field range
compared to its low critical temperature superconductor (LTS)
counterpart.

In this respect it is of particular importance to understand the
microscopic properties and dynamics of charge transport across
nano-sized GB JJs. Here, the investigation of low frequency electric
noise is a very useful tool to study the dynamics of both cooper
pair and quasiparticle charge transport, revealing among other
things information about the nature of the GB interface and its
homogeneity. Still after numerous experimental studies on HTS GB
junctions during the last decades the underlying physical transport
mechanisms across the GB interface are subject of recurring
discussion.

A large number of noise studies have been performed on wide
bicrystal and biepitaxial GB JJs with junction widths ranging from
one to several tens of micrometers
\cite{Kawasaki1992,Miklich,Marx1995a,Marx1995b,Marx1998,Liatti2006}.
Only a few electric transport studies have been performed on
sub-micrometer bicrystal GBs, where single charge trapping states,
responsible for the low frequency fluctuations of the transparency
of the GB barrier, could be resolved \cite{Herbstritt}.

In this article we compare two methods to fabricate YBCO Josephson
Junctions at the nano scale and their respective noise properties.
These methods are based on biepitaxial grain boundaries created in
single layer YBCO films. Both a conventional technique, where the
nanosized junctions are patterned by electron beam lithography and
ion milling, and a new technique, where the junctions are formed as
a result of phase competition between superconducting and insulating
phases at the grain boundary interface, will be described. We have
previously shown that the two methods give Josephson junctions with
fundamentally different critical current density $j_C$ and
resistivity $\rho_N$ values\cite{Davidnanoletter}. In this paper we
compare the noise data of soft nanopatterned GB junctions to various
electrical transport models, which allows us to determine the nature
of the biepitaxial GB barriers. Moreover, from the analysis of
single charge trap states in the GB barriers we are able to
qualitatively and quantitatively assess the detrimental effect of
ion milling on GBs during the conventional fabrication.

This paper is organized as follows. In section \ref{sec:Sample
fabrication} we describe both the conventional and soft
nano-structuring of biepitaxial YBCO JJs. Section
\ref{sec:Comparison} is dedicated to the comparison of the dc
transport properties between junctions fabricated with the two
nano-patterning methods. The noise models applicable to GB JJs are
introduced in section \ref{sec:NoiseTheory}. In section
\ref{sec:Results} we present the results and discussion of noise
measurements on JJs fabricated with the two nano-lithographic
methods.

\section{Sample fabrication}
\label{sec:Sample fabrication}
\subsection{Conventional nanostructuring}
The conventional way to fabricate deep submicron biepitaxial
Josephson junctions is to use electron beam lithography in
combination with a hard mask and ion beam milling. This procedure,
with amorphous carbon as hard mask, is well established and has been
proven to work well for the realization of various kinds of
submicron HTS Josephson junctions, for example ramp
type\cite{Komissinski}, bicrystal\cite{tobias} and
biepitaxial\cite{DaniellaJAP}. In this work we have fabricatated
deep submicron Josephson junctions by the biepitaxial technique.
Details on the fabrication procedure can be found
elsewhere\cite{DaniellaJAP, DaniellaIEEE}. Here we only summarize
the main steps. First a 30 nm thick SrTiO$_3$ (STO) layer is
deposited on a MgO (110) substrate using Pulsed Laser Deposition
(PLD). Next an amorphous carbon mask is deposited and then patterned
using e-beam lithography and oxygen plasma. Part of the seed layer
is then removed using Ion milling. Then a 100-120 nm thick YBCO film
is grown by PLD at a temperature of 790$^\circ$C . The film will
grow (001) oriented on the MgO substrate and (103) on the STO seed
layer. The YBCO film is then patterned using ion milling through an
amorphous carbon mask defined by e-beam lithography. Even though
junctions with widths smaller than 100 nm can, in principle, be
fabricated with this procedure, the damage caused by the ion milling
process will effectively limit the smallest possible width. The
damaged grain boundary region on both sides of the junction
constitutes a significant part of the total junction width, which
strongly affects the superconducting properties. We have therefore
engineered an alternative way to nanostructure HTS Josephson
junctions, which is described in the following section.

\subsection{Soft nanostructuring}
\begin{figure}
 \includegraphics{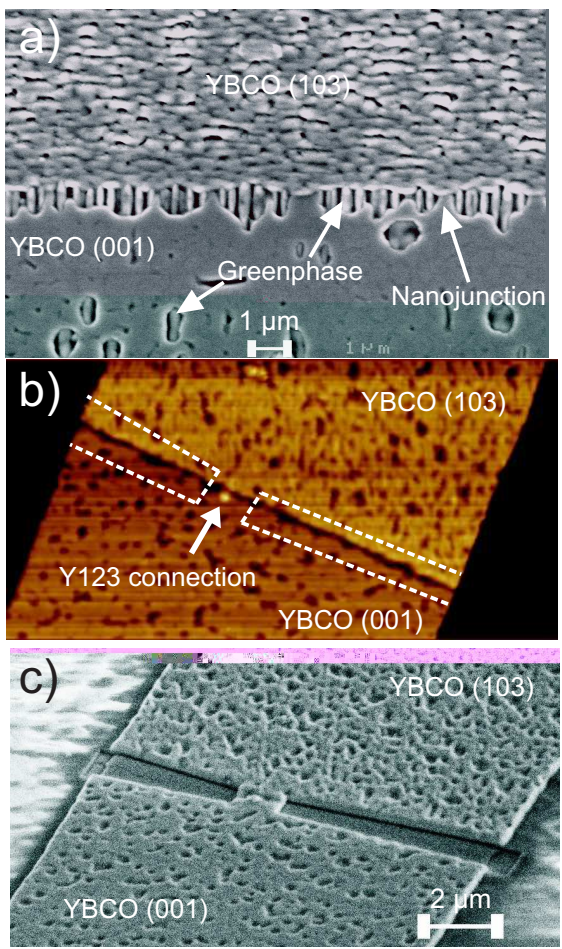}
  \caption{(a) SEM image of an interface between a (001) and (103) YBCO film. A significant amount of greenphase is present near the grain boundary.
  In a different study\cite{Davidnanoletter} transmission electron microscopy and energy dispersive x-ray analysis was used to
  confirm that the precipitate at the grain boundary is greenphase. (b) AFM scan of a 10 $\mu$m wide grain boundary interface before the FIB procedure.
  (c) SEM image of the same interface after the unwanted YBCO have been removed by the FIB leaving only one or two connections. }
  \label{Fig1}
\end{figure}

We have developed a new soft patterning method that allows
fabricating biepitaxial grain boundary junctions at the nanoscale
without significant lateral damaging effects due to the ion milling.
The procedure is based on the competition between the nucleation of
the superconducting and insulating phases at the grain boundary. To
fabricate the junctions we use the fact that for certain deposition
conditions secondary insulating phases like Y$_2$BaCuO$_5$ (Y211,
also called greenphase) can nucleate on MgO(110) in addition to the
superconducting YBa$_2$Cu$_3$O$_{7-\delta}$ (Y123). The amount of
greenphase increases for non optimal deposition conditions and in
the presence of grain
boundaries\cite{Green,Gustafsson2010,DiChiara}. Nano sized
superconducting Y123 connections embedded in a greenphase matrix are
expected to be formed at the grain boundary, see Figure \ref{Fig1}
(a). These connections can be isolated using a Focused Ion Beam
(FIB), see Figure \ref{Fig1} (b) and (c).

We first fabricate 10 $\mu$m wide grain boundary junctions using the
conventional method. A deposition temperature of 740$^\circ$C is
used for the YBCO film. The grain boundary is then examined using
atomic force microscopy (AFM) and scanning electron microscopy
(SEM). A suitable superconductive nanoconnection is selected and
then isolated by using FIB. By leaving greenphase regions of at
least 300 nm on each side of the Y123 connection, nanosized
Josephson junction with no lateral damage are created since the Ga
ions will only get implanted into the greenphase layer. Figure
\ref{Fig1} (c) shows a final device, where we have isolated an
approximately 200 nm wide junction protected on both sides by
greenphase.

\section{Comparison between the transport properties of conventional and soft nanopatterned junctions}
\label{sec:Comparison} 
\begin{figure}
 \includegraphics{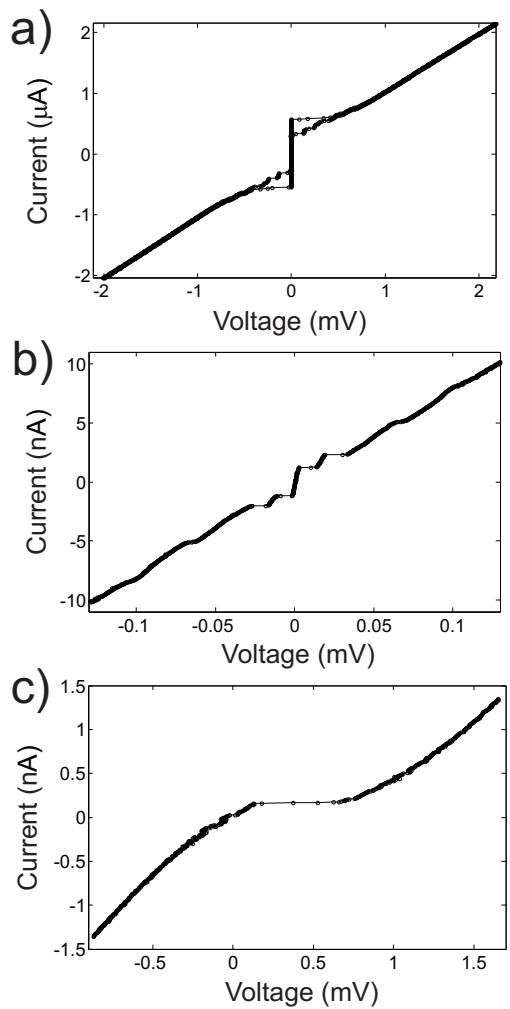}
  \caption{Current voltage characteristics for (a) A nanojunction fabricated using the
soft nanopatterning technique. The width of the junction extracted
from AFM is 200 nm. (b) a sample fabricated with the conventional
nanopatterning technique. The second switch is because this specific
sample was designed to have 2 Josephson junctions in series to allow
study of charging effects. Here the nominal junction width is 300
nm. (c) A sample fabricated using the conventional method, 200 nm
wide, with a coulomb blockade like behavior and no critical current.
The measurements were done at 271, 16 and 22 mK, respectively.}
  \label{Fig2}
\end{figure}

Electrical properties such as critical
current density ($j_C$), specific resistance ($\rho_N$) and critical
current ($I_C$) vs magnetic field ($B$) have been extensively
examined\cite{Davidnanoletter} and have shown significant
differences for the two fabrication methods. Figure \ref{Fig2} shows
the current voltage characteristics (IVC) for (a) a soft
nanopatterned junction (200 nm wide), (b) a 300 nm wide
conventionally patterned junction and (c) a 200 nm wide
conventionally patterned junction. A recurring pattern is seen here:
the soft nanostructured junctions have an order of magnitude or more
higher $j_C$ and one or several orders of magnitude lower $\rho_N$
when compared to conventionally fabricated samples. Conventionally
fabricated junctions with a width of 200 nm or less have high
resistive nonlinear IVCs with a suppressed Josephson current. Only
junctions with widths 300 nm or more showed a Josephson current.

\begin{figure}
 \includegraphics{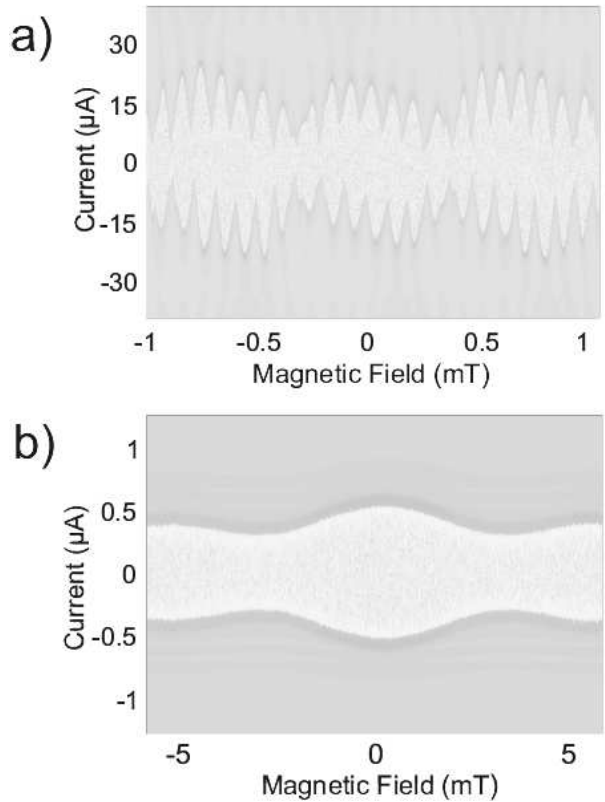}
  \caption{I vs B for (a) a 10 $\mu$m wide grain boundary with many parallel channels.
  (b) a sample cut by FIB, which consists of only 1 or 2 parallel channels.
  The grey scale represents the logarithmic conductance and the darkest region corresponds to $I_C$.}
  \label{Fig3}
\end{figure}

$I_C$ vs $B$ measurements revealed significant differences in the
modulation period for the two fabrication methods. The period of the
magnetic field modulation ($\Delta B$) of the Josephson current can
be used to approximate the width ($w$) of the region exhibiting
Josephson coupling in the junctions. Figure \ref{Fig3} (a) shows the
$I_C$ vs. $B$ for a 10 $\mu$m wide grain boundary junction before
the FIB cut to isolate the nanojunction; The behavior of the
magnetic pattern is that of several parallel
junctions\cite{Tinkham,Likharev}. Figure \ref{Fig3} (b) shows the
magnetic pattern of a softpatterned junction after the FIB cut,
leaving only one or two connections.

Depending on the electrode geometry we use the two expressions for
$\Delta$B as a function of the junction width $w_j$ from Rosenthal
and coworkers\footnote{The two formulas for $\Delta B$ are valid in
the short junction limit, $w_j$<4$\lambda_j$\cite{Likharev}, where
$\lambda_j$ is the Josephson penetration depth\cite{VanDuzer}:
$\lambda_j^2=\hbar/(2ej_C\mu_0(\lambda_{001}+\lambda_{103}+d))$. For
the soft nanopatterned junctions we get $\lambda_j \simeq$
1.2$\mu$m. All measured junctions are in the short junction limit
since $w_j$<4$\lambda_j$.}\cite{Rosenthal}: For the soft patterned
junctions having wide electrodes $w_e\simeq 10\mu$m we use the thick
electrode limit expression
\begin{equation}\label{Rosenthal1}
\Delta B=\frac{\Phi_0t}{1.2w_j^2(\lambda_{103}+\lambda_{001} +d)}
\end{equation}
valid for $\lambda_{001,103}^2/t < w_e$. Here $\lambda_{001}$ and
$\lambda_{103}$ are the London penetration depth in the (001) and
(103) electrode, respectively. $\Phi_0$ is the magnetic flux
quantum, $t$ is the thickness of the film and $d$ is the thickness
of the junction barrier. For the conventionally nanopatterned
junctions the width of the electrodes is equal to the width of the
junctions $w_e \simeq 300-500$ nm. Here the thin electrode limit
($\lambda_{001,103}^2/t \geq w_e$) applies
\begin{equation}\label{Rosenthal2}
\Delta B=\frac{1.84\Phi_0}{w_j^2}.
\end{equation}
The London penetration depth in the (001) electrode is given by the
penetration depth in the ab-planes $\lambda_{001} = \lambda_{ab}$.
Instead, as a result of the London penetration depth anisotropy in
YBCO $\lambda_{103}$ is given by a combination of $\lambda_{ab}$ and
the c-axis penetration depth $\lambda_c$, which depends on the grain
boundary angle\cite{Jesper}\footnote{$\lambda_{ab}$ for a critical
temperature (T$_c$) of 89 K is approximately 160 nm\cite{Zuev}. The
effective London penetration depth, $\lambda_{eff}$,  was obtained
using $\lambda_{c}$ = 2 $\mu$m\cite{Homes} and geometrical
considerations\cite{Jesper} with a dependence on the grain boundary
angle.}.

Equation \ref{Rosenthal1} was used on a number of soft patterned
junctions and the extracted width was compared to the nominal width
measured by AFM and SEM\cite{Davidnanoletter}. The values were at
most differing by 40\%; This shows that the width of the
superconducting transport channels extracted from the magnetic
pattern was very close to the measured junction widths.

For one of the conventionally nanopatterned junction a $\Delta B$ of approximately 1 T was extracted from the magnetic pattern.
 Using equation \ref{Rosenthal2} resulted in a width of 60 nm, significantly less than the nominal junction width of 300 nm.
 Similar results where obtained for two other junctions 300 and 500 nm wide.
This in combination with the j$_C$ and $\rho_N$ values shows that a
substantial part of the grain boundary, approximately 100 nm wide,
on each lateral side of the junction does not feature any Josephson
coupling.

The magnetic patterns of conventionally nanopatterned junctions have revealed the presence of a highly non uniform grain boundary,
having a much reduced region with Josephson coupling compared to the nominal one.
However, this does not give a clear image of the total area which retains the Josephson coupling along the grain boundary. In fact, it only tells
us that the largest spacing between superconducting channels is significantly less than the nominal junction width.
To estimate the area of both the Cooper pair and quasiparticle transport channels we analyzed the voltage noise of the junctions caused by single charge traps in the GB barrier, which will be discussed in section \ref{sec:Results}.

\section{Noise theory for grain boundary junctions}
\label{sec:NoiseTheory}

Noise measurements are a helpful tool to extract
information about the electrical transport through the junction and
hence to obtain information about the nanostructure of the grain
boundary interface. In this work we focused on the low frequency
noise spectra of both the critical current fluctuations $\delta I_C$
and normal resistance fluctuations $\delta R_N$, which are related
to the transport mechanisms of the cooper pairs and quasiparticles,
respectively.

It is well established that at low frequencies the critical current
and normal resistance fluctuations are governed by bistable charge
trapping states in the junction barrier \cite{Rogers}. The trapping
of a charge will locally increase the junction barrier making it
less transparent. This process can be considered as a reduction of
the total junction area $A_{j}$ by an amount which is
proportional to the cross section of the localized charge trap
state $A_{t}$. The fluctuating barrier transparency (or equivalently
junction area) results in fluctuations of the critical current $I_C$
and normal resistance $R_N$. Each individual charge trap causes a
random telegraph switching (RTS) signal between two states, with
respective mean lifetimes $\tau_1$ and $\tau_2$,
of both the junction normal
resistance and critical current. The corresponding frequency spectrum is given
by a Lorentzian
\cite{Machlup}:
\begin{eqnarray}\label{Lorentzian}
  S_R^{RTS}(f) & = & \frac{4\langle(\frac{\delta R_N}{R_N})^2\rangle \tau_{eff}}{1+(2\pi f \tau_{eff})^2}, \nonumber\\
  S_I^{RTS}(f) & = & \frac{4\langle(\frac{\delta I_C}{I_C})^2\rangle \tau_{eff}}{1+(2\pi f
  \tau_{eff})^2},
\end{eqnarray}
where $\tau_{eff} = (\tau_1^{-1}+\tau_2^{-1})^{-1}$ is the effective
lifetime of the underlying RTS signal and $f$ is the frequency.
$\langle(\delta R_N /R_N)^2\rangle$ and $\langle(\delta I_C
/I_C)^2\rangle$ are the mean squared relative fluctuations of the
normal resistance and critical current caused by the charge trap.
For large enough junction areas many bistable charge trapping states
will contribute to the total noise. Assuming a constant distribution
of transition rates $1/\tau_{eff}$ the resulting noise power
spectrum will have a $1/f$ shape.

The values of the relative root mean square (rms) fluctuations
$\delta I_C/I_C$ and $\delta R_N/R_N$ can be determined by measuring
the voltage noise across the junction at various bias current
values. For a Josephson junction having a non hysteretic current
voltage characteristic the total voltage fluctuations across the
junction at a fixed bias current $I$ are given by\cite{Miklich}
\begin{equation}\label{SV}
  S_V(f)=(V-R_dI)^2S_I(f)+V^2S_R(f)+k(V-R_dI)VS_{IR}(f),
\end{equation}
where $V$ is the dc-voltage across the junction, $R_d = \partial
V/\partial I$ is the differential resistance, $S_I =|\delta
I_C/I_C|^2$, $S_R =|\delta R_N/R_N|^2$, and $S_{IR} =|\delta
I_C/I_C||\delta R_N/R_N|$ is the cross spectral density of the
fluctuations. Here it is assumed that $S_I$ and $S_R$ are composed of an ensemble of
RTS signals, $S_I^{RTS}$ and $S_R^{RTS}$, respectively. The value
$k$ represents the correlation between the $\delta I_C$ and $\delta
R_N$ fluctuations. One has $k=-2$ and $k=2$ for perfectly antiphase
and inphase correlated fluctuations, respectively. For uncorrelated
fluctuations one obtains $k=0$. From equation \ref{SV} it follows
that at bias currents close to the critical current the voltage
fluctuations are dominated by critical current fluctuations $S_I$
due to the large differential resistance. For large bias currents,
where the differential resistance approaches the asymptotic normal
resistance, the voltage noise is governed by resistance fluctuations
$S_V = V^2S_R$. The correlation term $S_{IR}$ will only contribute
to the voltage noise in the intermediate bias current regime, while
it is negligible close to the critical current and for large bias
currents.

The values of the relative fluctuations $\delta I_C/I_C$ and $\delta
R_N/R_N$ depend on the nature of the junction barrier. Indeed, from
the ratio $q=|\delta I_C/I_C|/|\delta R_N/R_N|$ between the relative
fluctuations one can extract information about the homogeneity of
the junction barrier as we will discuss on the basis of the
following three junction models applicable to grain boundary
junctions:

For a homogenous junction barrier one can assume that the $I_CR_N$
product is a constant, independent of the critical current density
$j_c$ and resistivity $\rho_n$. This is for example the case for a
superconductor-insulator-superconductor (SIS) junction
\cite{Tinkham}, where cooper pairs and quasiparticles tunnel
(directly) through the same parts of the junction.
From the constant $I_CR_N$ product it follows
directly that the relative fluctuations of the critical current and
normal resistance have the same amplitude and are anticorrelated
$\delta I_C/I_C = -\delta R_N/R_N$, resulting in a ratio $q=1$.

In the Intrinsic Shunted Junction (ISJ) model\cite{Gross,GrossPRB},
instead, where the barrier is assumed to be inhomogeneous
containing a high density of localized electronlike states, the
quasiparticle transport is dominated by resonant tunneling
via the localized states.
On the contrary, due to Coulomb repulsion cooper pairs can only tunnel
directly through the barrier. Detailed calculations show that the
$I_CR_N$ product is not anymore constant, instead it follows the
scaling behavior $I_C R_N \propto (j_c)^p$, where $j_c$ is the
critical current density of the junction and $p$ is a constant
depending on the position of the localized states. For localized
states sitting in the middle of the barrier the scaling power is
$p=0.5$. From this $I_CR_N$ scaling behavior one obtains for the
ratio of the normalized fluctuations $|\delta I_C/I_C|/|\delta
R_N/R_N|=q=1/(1-p)$\cite{Gross}. Typical experimental values of
$q$ range between 2 and 4
\cite{Kawasaki1992,Marx1995a,Marx1995b,Marx1998}.

The channel model proposed by Micklich {\it et al.} \cite{Miklich}
assumes a junction which consists of N parallel channels, where all
channels have the same resistance but only one channel carries a
supercurrent. For a large number of channels $N$ the fluctuations in
critical current can be much higher than the fluctuations in
resistance, giving a high ratio $q$. Since the supercurrent and the
quasiparticles have separate channels no correlation between the
critical current and resistance fluctuations is expected.

\begin{figure}
 \includegraphics[width=90mm]{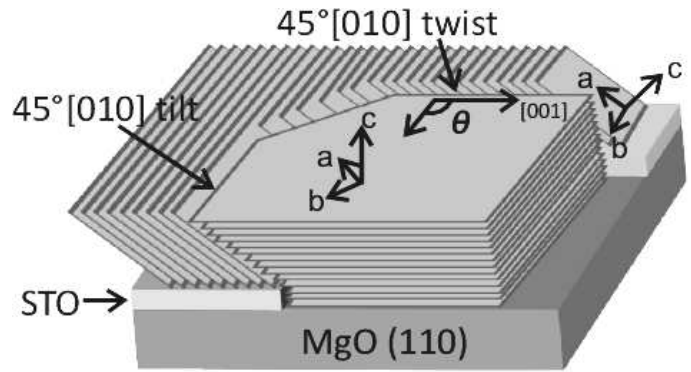}
  \caption{a) Sketch of the interface geometry. The crystallographic orientations of the (001) and (103)
  YBCO is indicated by arrows. $\theta$ is the interface angle and is defined with respect to the [001] MgO direction.}
  \label{Fig8}
\end{figure}

\begin{figure}
 \includegraphics[width=90mm]{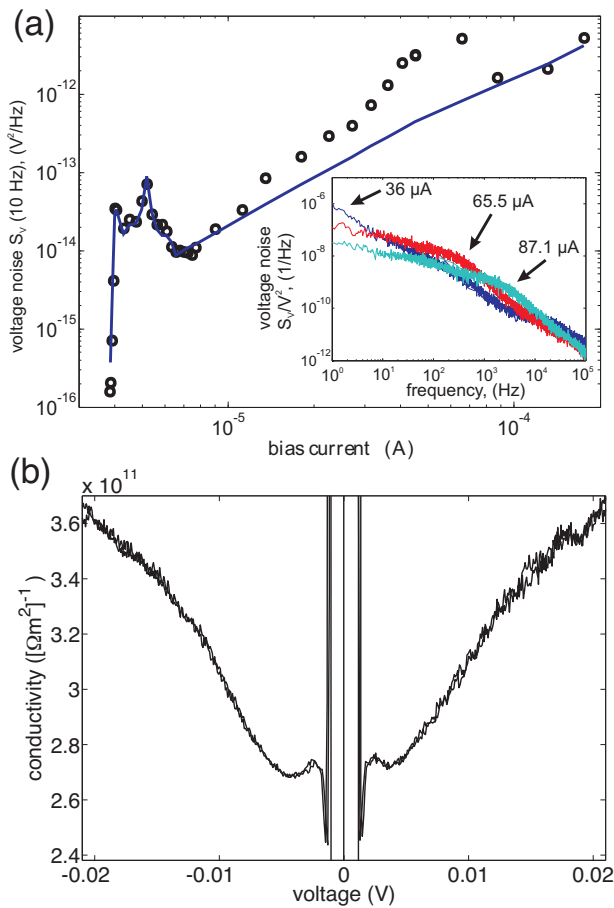}
  \caption{(a) Voltage noise at 10 Hz as a function of bias
current (open symbols) for a soft nano patterned junction (nominal
interface angle = 30$^\circ$). A theoretical fit (solid line) is
included to determine S$_I$ and S$_R$. The inset shows the voltage
noise as a function of frequency for three different bias points.
The "hump" moves to higher frequencies when the bias is increased.
(b) Conductivity as a function of bias voltage of the soft nano
patterned junction. The central part ($|V|<1$) mV is related to the
Josephson effect.}
  \label{Fig4}
\end{figure}

\begin{figure}
 \includegraphics[width=90mm]{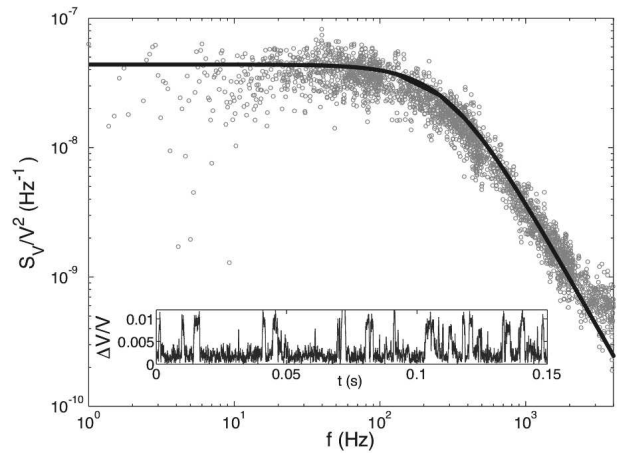}
  \caption{Noise spectrum measured at $I \gg I_C$
  after the subtraction of the $1/f$ background. The black line
  is a Lorentzian fit to equation \ref{SV_RTS}. The plateau of the Lorentzian
  is given by $4 \tau_{eff}\langle (\delta V/V)^2 \rangle$, with
  $2 \pi \tau_{eff} = 302$ Hz and $\delta V/V = 0.0044$. The inset
  shows the respective time trace with $\Delta V/V \simeq 0.0095$.}
  \label{Fig5}
\end{figure}

\begin{figure}
 \includegraphics[width=90mm]{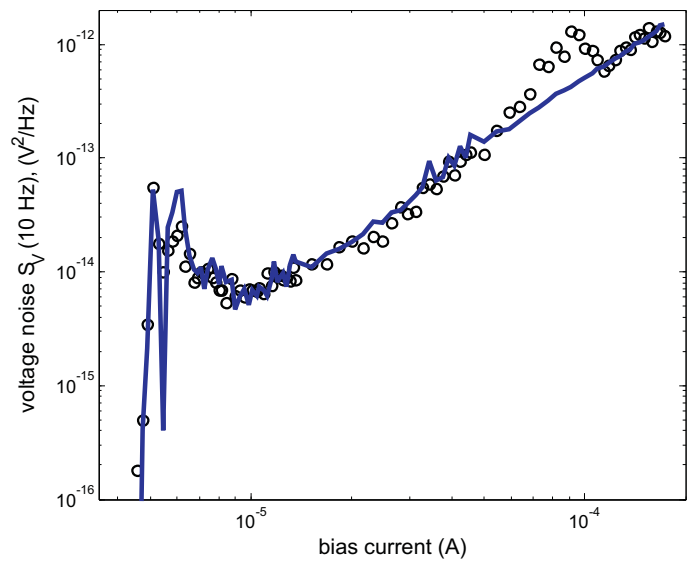}
  \caption {Voltage noise at 10 Hz as a function of bias
current (open symbols) of a soft nano patterned junction with an
interface angle of 50$^\circ$. The solid line is a theoretical fit
used to determine S$_I$ and S$_R$.}
  \label{Fig10}
\end{figure}

\section{Results and Discussion}
\label{sec:Results}

The voltage noise spectral density was measured using a room
temperature voltage preamplifier with an input noise of 4 nV/Hz$^{1/2}$ followed by a Stanford
Research Dynamic Signal analyzer SR785 for a number of bias points.
These measurements were done at 4 K for the soft nanostructured
junctions, where the current voltage characteristic was non
hysteretic. For the conventionally fabricated samples a temperature
of 280 mK was needed to avoid thermal smearing of the IVC
(due to the low $I_C$ of these junctions)\cite{Miklich}.

All noise measurements on the conventionally patterned samples refer
to grain boundaries obtained by patterning the STO seed layer
parallel to the [100] direction ($\theta =  0^\circ$) of the MgO
substrate\cite{Jesper} (see Fig. \ref{Fig8}). The soft patterned
junctions had nominal interface angles ($\theta$) of 30$^\circ$ and
50$^\circ$. Grain boundaries with a small interface angle $\theta$
have a micro structure close to a basal-plane like type ($45^\circ$
[010] tilt), see Fig. \ref{Fig8}, where the a-b planes of the (001)
YBCO electrode meet one single a-b plane on the (103) YBCO electrode
side\cite{Cedergren}. These grain boundaries have proven to be low
dissipative \cite{Bauch2005,Bauch2006,Lombardi_PhysicaC} compared to
[001]-tilt ones and suitable for applications in quantum circuitry.
By increasing the GB angle $\theta$ the interface gradually evolves
into a $45^\circ$ [010] twist GB at $\theta = 90^\circ$ (see Fig.
\ref{Fig8}).

\begin{table*}
\caption{\label{Table_coll} Collection of 1/f noise in HTS GB Josephson junctions of different
HTS material, GB interface type, technology, and junction area $A_j$. The interface angle $\theta$
of the soft patterned junctions is illustrated
in Fig. \ref{Fig8}. The values $\alpha$ and $\beta$ are given by $\cos^2(2\pi\theta/360)$
and $\sin^2(2\pi\theta/360)$, respectively.}
\begin{ruledtabular}
\begin{tabular}{cccccc}
Junction & GB & $A_{j}$ & $S_I^{1/2}$ (10 Hz)  & $q=(S_{I}/S_{R})^{1/2}$ & $A_{j}^{1/2}S_{I}^{1/2}$ (10 Hz)\\
technology & type & ($\mu$m$^2$) & $\times 10^{-4}$(Hz$^{-1/2}$) & & $\times 10^{-6}(\mu$m/Hz$^{1/2})$ \\
\hline
    YBCO/STO/MgO biepitaxial &  $45^\circ$ [010] tilt & 0.02 & $1.0\pm 0.1$ & $1.0\pm 0.1$ & 14 \\
     $\theta\simeq 0^\circ$ (this work) & &  & ($T=4.2$K) &  &  \\
        &  &  &  &  &  \\
     YBCO/STO/MgO biepitaxial  & $\alpha \cdot \{45^\circ$ [010] tilt$\}+$ & 0.052 &  $1.6\pm 0.1$ & $1.8\pm 0.2$ & 36 \\
      $\theta\simeq 50^\circ$ (this work) & $\beta \cdot \{45^\circ$ [010] twist$\}$ &  & ($T=4.2$K) &  &  \\
         &  &  &  &  &  \\
      YBCO/NdGaO$_3$ bicrystal  & $2\times14^\circ$ [100]  tilt & 0.06 & $0.36\pm 0.06$  & $1.05\pm 0.1$ & 10 \\
     (Ref. \cite{Liatti2006}) &  &  & ($T=55$K) &  &  \\
        &  &  &  &  &  \\
     YBCO/STO bicrystal  & $25^\circ$ [001]  tilt & 1.0 & $0.32$  & $2.5$ & 32 \\
     (Ref. \cite{Kawasaki1992}) &  &  & ($T=70$K) &  &  \\
        &  &  &  &  &  \\
    Bi$_2$Sr$_2$CaCu$_2$O$_{8+x}$/STO bicrystal  & $24^\circ$ [001]  tilt & 1.6 & $0.24\pm 0.06$  & $1.9\pm 0.3$ & 30 \\
     (Ref. \cite{Marx1995a}) &  &  & ($T=40-70$K) &  &  \\
        &  &  &  &  &  \\
    YBCO/STO bicrystal  & $24^\circ$ [001]  tilt & 3.8 & $0.18\pm 0.01$  & $3.8\pm 0.6$ & 35 \\
     (Ref. \cite{Marx1995b,Marx1998}) &  &  & ($T=25-70$K) &  &  \\
        &  &  &  &  &  \\
     YBCO/STO bicrystal  & $36.8^\circ$ [001]  tilt & 1.0 & $0.35\pm 0.12$  & $3.7\pm 0.8$ & 35 \\
     (Ref. \cite{Marx1995b,Marx1998}) &  &  & ($T=30-70$K) &  &  \\
        &  &  &  &  &  \\

\end{tabular}
\end{ruledtabular}
\end{table*}

\subsection{Noise properties of soft patterned junctions}
In Figure \ref{Fig4} (a) the voltage noise spectral density at 10 Hz
is plotted as a function of the bias current $I$ for one of our soft
patterned nanojunctions having a width $w\simeq 200$ nm and nominal
interface angle of 30$^\circ$. However, by examining the grain
boundary by AFM the actual interface angle was closer to 0$^\circ$.

As expected, the noise peaks close to the critical current when
$I\simeq I_C = $4 $\mu$A. A second peak appears close to $I=$5.2
$\mu$A. This is due to a resonance feature in the IVC causing the
differential resistance (R$_d$) to spike. For higher bias, where the
resistance fluctuations dominate, the noise increases quadratically.
The hump structure around 50 $\mu$A is caused by a single charge
trap causing a RTS signal with a typical Lorentzian spectrum on top
of a $1/f$ background. The occurrence of such Loretzians is typical
for a limited number of charge traps in submicron sized GB junctions
\cite{Herbstritt}. The voltage dependence of the effective lifetime
of the charge trap causes the Lorentzian to move to higher
frequencies for increasing bias current, which is shown in the inset
of Figure \ref{Fig4}.

To fit the measured voltage noise spectral density at 10 Hz as a
function of bias current (see solid line in Figure \ref{Fig4} (a))
to equation \ref{SV}, which assumes a pure $1/f$ noise spectrum, we
neglected the data between 20 $\mu$A and 80 $\mu$A caused by a
single charge trap. From the fit we obtain $S_{I}\simeq S_R\simeq
10^{-8}/$Hz resulting in $q=|\delta I_C/I_C|/|\delta R_N/R_N|\simeq
1$, and $k\simeq -1.3$. The ratio $q\simeq 1$ indicates that our
junction has a rather homogeneous barrier, where quasi particles and
Cooper pairs tunnel directly through the same parts of the barrier
\cite{Miklich}. Together with a tunnel like conductance spectrum
(see Figure \ref{Fig4} (b)) we can conclude that our junction
barrier is very similar to that of a SIS junction, consistent with
the band bending model \cite{Mannhart1998,Hilgenkamp1998}. Similar
results have only been found in $2\times 14^{\circ}$ [100]-tilt YBCO
GB Josephson junctions \cite{Liatti2006}. Furthermore, our result is
incompatible with the Intrinsically shunted junction model
\cite{Gross,GrossPRB,Marx1995a,Marx1995b}
and the channel model \cite{Miklich},
where $q$-values larger than 2 are expected.
The deviation of the correlation between the critical current and
resistance fluctuations from perfect anti-correlation ($k = -2$)
could be caused by the limited amount of two level fluctuators in
the small junction area not representing a perfect ensemble. It is
important to point out that the noise properties of our nano
junction close to an ideal SIS Josephson junction underline once
more the pristine character of the junction barrier that can be
obtained by using the soft nano-patterning method.

In Figure \ref{Fig10} the voltage noise spectral density at 10 Hz is
plotted for a 520 nm wide sample having a nominal interface angle of
$\theta\simeq50^\circ$. This value was confirmed by the AFM
inspection of the GB. From the fit we obtain $S_{I}\simeq 2
 \cdot10^{-8} - 3 \cdot10^{-8} /$Hz, $S_R\simeq 8 \cdot 10^{-9}/$Hz, and $k\simeq -0.5$ resulting in $q=1.8 \pm0.2$.
The $q$ value close to 2 indicates that the transport across the GB
barrier cannot be described by a direct tunneling model, e.g. a
homogeneous SIS tunnel junction. Instead, our result shows that for
this kind of GB type (mixture of $45^\circ$ [010] twist and
$45^\circ$ [010] tilt) the barrier is better described by the ISJ
model, where
 quasiparticles tunnel resonantly via localized states.

In table \ref{Table_coll} we summarize the noise data of the soft
nanopatterned biepitaxial YBCO GB junctions together with results
from literature on HTS Josephson junctions of various GB types.
Comparing the ratios $q=(S_{I}/S_{R})^{1/2}$ between different GB
types one can clearly see that only GBs where the ab-planes in at
least one of the electrodes are tilted around an axis parallel to
the GB interface, e.g. $45^\circ$ [010] tilt (this work) and
$2\times 14^\circ$ [100] tilt \cite{Liatti2006}, have a ratio
$q\simeq 1$. All the other GB types such as [001] tilt
\cite{Kawasaki1992,Marx1995a,Marx1995b,Marx1998} and $\alpha \cdot
\{45^\circ$ [010] tilt$\}+\beta \cdot \{45^\circ$ [010] twist$\}$,
with $0<\beta \leq 1$ exhibit ratios $q\gtrsim 2$. These facts give
a strong indication that the nature of HTS GB barriers depends on
how the ab-planes meet at the interface. GBs with ab-planes tilted
around an axis parallel to the GB interface, such as a basal plane
GB, can be described by a direct tunneling model consistent with a
homogeneous SIS barrier. All other GB types deviating from a bare
rotation of the ab-planes around the GB line are characterized by
resonant quasiparticle tunneling via localized states (ISJ model).

From the spectral density of the critical current fluctuations and
the junction area one can obtain information about the areal charge
trap density $n_t$ and the cross sectional area $A_t$ of the charge
traps \cite{Marx1997,VanHarlingen2004}. Assuming $N$ identical and
independent charge traps the spectral density of the relative
critical current fluctuations scales with the junction area $A_j$ as
$\langle (\delta I_C/I_C)^2 \rangle = N(A_t/A_j)^2 = n_t A_t^2/A_j$.
From this equation it follows that the quantity $(S_I A_j)^{1/2}$ is
proportional to the product of cross sectional area of a charge trap
and the square root of the trap density $A_t n_t^{1/2}$. In table
\ref{Table_coll} we show the computed product $(S_I A_j)^{1/2}$ at
10 Hz for the various GB types. Remarkably the values for GBs having
ratios $q\geq2$ are close to $35\times 10^{-6} \mu$m/Hz$^{1/2}$
\footnote{The constant value $(S_I A_j)^{1/2}$ is equivalent to the
scaling behavior of the resistance fluctuations $S_R \propto R_N$
reported by Marx et al. in Ref. \cite{Marx1997}.}. Instead, the
values for $(S_I A_j)^{1/2}$ in GB types with $q\simeq 1$ are
roughly 3 times smaller. Assuming that the cross sectional area of a
charge trap is independent of the GB type, the difference in charge
trap density supports once more the different nature of the GB
barriers.

In the following we will use the Lorentzian spectra sitting on top
of a $1/f$ background (see inset of Fig. \ref{Fig4}) to estimate the
cross sectional area $A_{t}$ of a single charge trap in the barrier:
A single charge trap causes the voltage across the junction to
fluctuate between two bistable states with an amplitude $\Delta V$
(see inset in Fig. \ref{Fig5}). For large bias currents $I \gg I_C$,
when the differential resistance is asymptotically reaching the
normal state resistance of the junction, we can write for the
respective relative resistance change $\Delta R_N/R_N =  \Delta
V/V$. Assuming that the current flow across the junction is
homogeneous and the charge trap completely blocks the current flow
in a small part of the junction barrier we can determine the charge
trap's cross sectional area $A_t$ from the measured voltage
fluctuation amplitude $\Delta V$
\begin{equation}\label{trap_area}
  A_t = \frac{\Delta R_N}{R_N} A_{qp} =  \frac{\Delta V}{V} A_{qp},
\end{equation}
where $A_{qp}$ is the total area of quasi particle transport along
the junction. Instead of extracting $\Delta V$ from a voltage time
trace, one can also use the mean squared fluctuation amplitude
$\langle (\delta V)^2 \rangle$ determined from a Lorentzian fit of
the noise spectrum (see Fig. \ref{Fig5}). The two quantities are
related via \cite{Herbstritt}
\begin{equation}\label{DeltaVtime}
  (\Delta V)^2 = \left( \frac{\tau_1}{\tau_2}+\frac{\tau_2}{\tau_1}+2\right) \langle (\delta V)^2 \rangle.
\end{equation}
For clearly visible Lorentzians in the measured noise spectra the
ratio between the two mean life times is typically in the range from
1 to 10. Hence, we can approximate the fluctuation amplitude within
a factor of two by $\Delta V \simeq 2 \sqrt{\langle(\delta
V)^2\rangle} $ using the root mean squared (rms) fluctuation
amplitude extracted from a Lorentzian spectrum (see Fig.
\ref{Fig5}):
\begin{equation}\label{SV_RTS}
  S_V^{RTS}(f)=V^2S_R^{RTS}(f).
\end{equation}

\begin{figure}
 \includegraphics{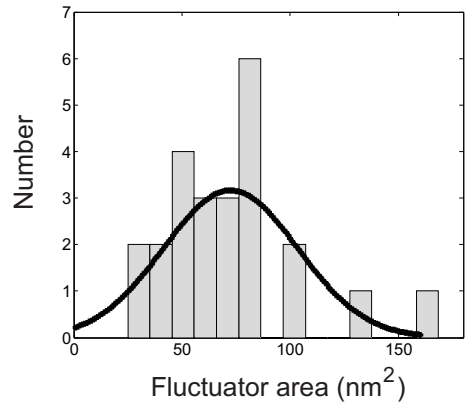}
  \caption{Histogram showing the effective fluctuator area that was extracted from multiple spectra of 3
   different soft patterned junctions in the high bias range. The black curve is a normal distribution with the same mean and standard deviation as the data set.}
  \label{Fig6}
\end{figure}

Together with equation \ref{Lorentzian} we can extract $\delta
R_N/R_N$ and approximate the cross sectional area of a charge trap:
\begin{equation}\label{DefectArea}
  A_t \simeq 2\frac{\delta R_N}{R_N} A_{qp}.
\end{equation}
 From the results of the previous section we can make the following considerations:
\begin{enumerate}
\item The $I_C$ vs $B$ measurements have shown that the
modulation period corresponds to an effective width close to the nominal
width of the junctions. We can therefore assume that the Cooper pair
transport is along the whole grain boundary,
$A_{cp}$ $\simeq$ $A_{j}$.
\item The fitting of the voltage noise spectral density has shown
that $S_{I}\simeq S_R$, this tells
us that the area of the superconducting channel is approximately equal to the
quasiparticle one, $A_{cp} \simeq A_{qp}$.
\end{enumerate}
These two facts imply that the areas of both transport channels are
very close to the nominal junction area ($A_{qp} \simeq A_{cp}
\simeq A_{j}$). One can, therefore, use the nominal area, measured
by AFM or SEM, in combination with the noise measurement to extract
$A_{t}$. Here we also use that the junction thickness $\simeq$ film
thickness (120 nm). We have made this type of analysis for 3 soft
patterned junctions and fitted a total of 24 Lorentzians in the high
bias range on different spectra. The extracted distribution of
$A_{t}$ are plotted in Figure \ref{Fig6}: We get an average area for
the fluctuators of about 72 nm$^2$, which is comparable to results
found in submicron [001]-tilt YBCO GB junctions \cite{Herbstritt}.

\begin{figure}
 \includegraphics{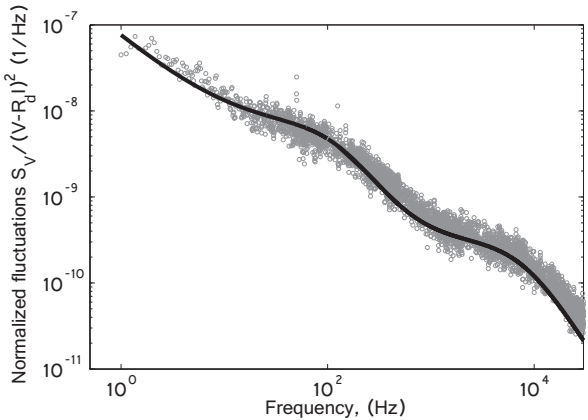}
  \caption{ Noise spectrum (open circles) and fit of two Lorentzians with a
  $1/f$-background (solid line) for one of the soft nanostructured
  junctions measured at $I\simeq I_C$.}
  \label{Fig7}
\end{figure}

In Figure \ref{Fig7} the spectrum at a bias current close to $I_C$ has been fitted by 2
Lorentzians and a weak 1/f background. Since the contribution from $R_N$
fluctuations is negligible for this range of currents one can extract $\delta
I_C/I_C$ using equation \ref{Lorentzian} and:
\begin{equation}\label{SI_RTS}
  S_I^{RTS}(f)=\frac{S_V^{RTS}(f)}{(V-R_dI)^2}
\end{equation}
The extracted values for $\delta I_C/I_C$ and $\delta R_N/R_N$ are
fairly similar in magnitude, $\delta I_C/I_C$ being at most 3 times
larger than the average of $\delta R_N/R_N$. This difference
could be explained by a spread in the fluctuators area. Indeed the values
of $\delta I_C/I_C$ extracted close to $I_C$ will certainly come from different two level
fluctuators than those generating $\delta R_N/R_N$ fluctuations at high biases.

\subsection{Noise properties of conventional junctions}
\begin{figure}
 \includegraphics[width=90mm]{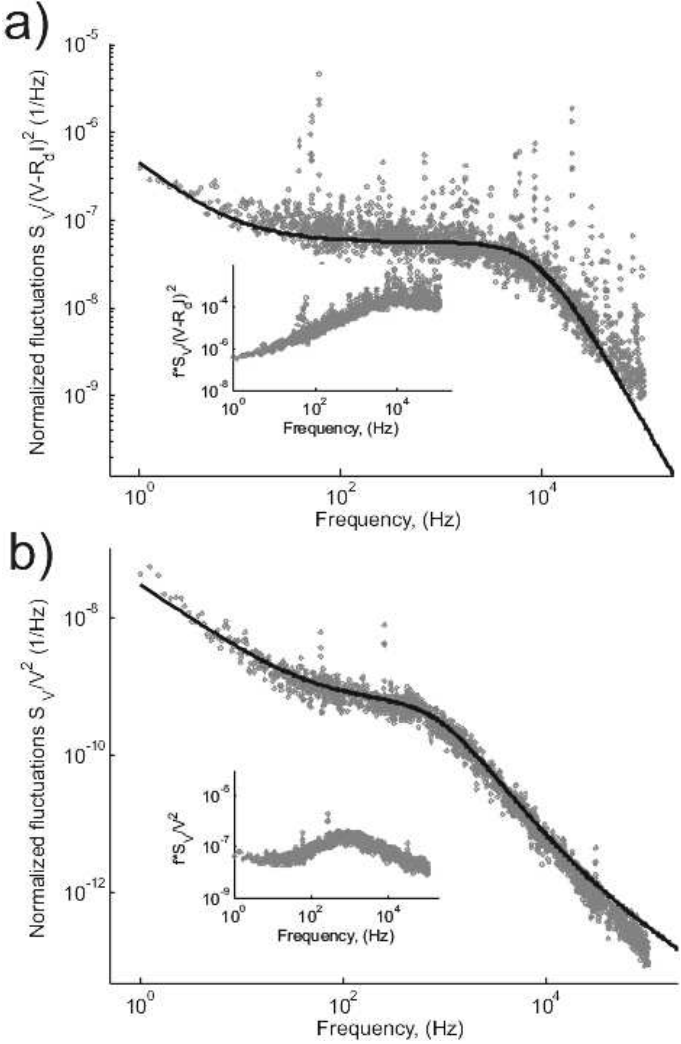}
  \caption{Noise spectrum (open circles) and fit of a Lorentzian with a
  $1/f$-background (solid line) for a conventionally patterned junction measured at a) $I\simeq I_C$ and b) $I \gg
  I_C$. The vertical axis show the normalized fluctuations
  corresponding to a) $S_I$ and b) $S_R$. The insets shows normalized fluctuations multiplied by the frequency to
  emphasize the amplitude of the charge traps.}
  \label{Fig9}
\end{figure}

Identical measurements and analysis were carried out for 3 nanosized
junctions fabricated by conventional nanolithography. For these
samples, at bias currents slightly above $I_C$ we have observed the
presence of strong Lorentzians in the low frequency spectra caused
by single charge traps, see Figure \ref{Fig9} (a). This circumstance
makes the fitting of the data to equation \ref{SV} in the low bias
range impossible, therefore preventing the extraction of $S_I$ and
the comparison with $S_R$. However, for these junctions we were able
to fit the Lorentzian voltage noise spectra for the two different
bias ranges, where they are dominated by current fluctuations (close
to $I_C$), $S_I^{RTS}$, and by resistance fluctuations (far above
$I_C$) , $S_R^{RTS}$, respectively, see Figure \ref{Fig9}. Assuming
that the average effective area of the charge traps is roughly the
same as that extracted from the junctions fabricated by soft
nanopatterning we can estimate A$_{cp}$ and A$_{qp}$ \footnote{The
$I_C$ vs B pattern of the conventional junctions do not follow a
Fraunhofer like pattern. This hints that the interfaces consists of
an array of superconducting channels. However, as long as the
individual channels have an area greater than the effective area of
the defects, equation \ref{DefectArea} is still valid.} for the
conventionally fabricated junctions. The part of the junction area
manifesting Josephson coupling and quasi particle transport can be
approximated by $A_{cp} \simeq A_t I_C/2\delta I_C$ and $A_{qp}
\simeq A_t R_N/2\delta R_N$, respectively. In the insets of Figure
\ref{Fig9} we show the spectral density of the normalized
fluctuations multiplied by the frequency. The pronounced difference
between the fluctuation amplitudes of the Lorentzians in the
critical current and resistance noise spectra by several orders of
magnitude clearly manifests the difference in area for the quasi
particle and cooper pair transport channels. In table \ref{Table1}
we summarize the results for 3 conventionally patterned junctions.
The average quasiparticle area is 25-50\% less than the nominal
area. However, the superconducting area varies greatly and for 2 of
the junctions it is significantly less than the quasiparticle area.
The A$_{qp}$ extracted from the noise measurements tells us that the
grain boundaries, despite losing most of the Josephson properties,
still have a quasiparticle transport channel with an average area
comparable to the nominal area. The area of the quasiparticle
channel does not decrease much in the fabrication process for the
conventional junctions, however the resistivity of the channel seem
to increase. Evaluating the critical current density based on the
effective area across which cooper pair transport occurs results in
$j_C^{eff} = I_C/A_{cp} \simeq 2-20$ kA/cm$^2$, where $I_C$ is the
critical current of the conventionally patterned junctions. Here it
is interesting to note that these values are in the same range as
those found in pristine soft nano-structured GB junctions
\cite{Davidnanoletter}. This fact clearly reflects the strong
dependence of the Josephson coupling on the stoichiometry in close
proximity (length scale of coherence length) of the GB. The
detrimental effect of the ion etching process during the
conventional nano patterning seems to locally kill the Josephson
coupling rather than inducing a gradual decrease over the whole
junction area. The ion beam procedure appears to be an on-off
process for the Josephson coupling: the grain boundary region which
survives the ion bombardment preserves the same Josephson properties
as the untouched soft nano-patterned samples. The overall increase
of more than one order of magnitude of the GB resistivity of
conventionally fabricated nanojunctions compared to the soft
nano-patterned ones can therefore be related to a reduced barrier
transparency in the junction regions where the Josephson coupling
has been switched off in the milling procedure.

\begin{table}
\caption{\label{Table1}Total nominal area $A_j$, area for the
quasiparticle transport channels $A_{qp}$ and  area for
superconducting transport channels $A_{cp}$ (extracted from noise
data) for 3 conventionally patterned samples.}
\begin{ruledtabular}
\begin{tabular}{cccc}
Junction nr & $A_{j}$ (nm$^2$) & $A_{qp}$ (nm$^2$) & $A_{cp}$ (nm$^2$)\\
\hline
    Nr 1 & 50000 & 31500 & 1250  \\
    Nr 2 & 30000 & 14600 & 9060 \\
    Nr 3 & 30000 & 22300 & 160 \\
\end{tabular}
\end{ruledtabular}
\end{table}

\section{Summary and Conclusions}
To summarize we have compared the electrical transport and noise
properties for nanoscaled biepitaxial YBCO grain boundary Josephson
junctions fabricated by two different methods. From electrical
transport and $1/f$ voltage noise properties of soft nanopatterned
Josephson junctions with a GB characterized by a rotation of the
ab-planes parallel to the interface (large c-axis tunneling
component) we can conclude that the GB barrier is very homogeneous
and has a SIS character (direct tunneling model). Instead, the noise
properties of soft patterned junctions, where the transport is
dominated by tunneling parallel to the ab-planes, are in accordance
with a resonant tunneling model (ISJ model). From the analysis of
two level fluctuators in the barrier, on the other hand, we find
that the conventional nanofabrication method severely deteriorates
the Josephson properties of the GB. The junction area maintaining
Josephson current can on average be much smaller than the nominal
area, while the quasi particle transport area is similar to the
nominal one. In this case the transport across the GB interface can
be well described by the transport model proposed by Miklich
\cite{Miklich}. The resistivity in these samples is increased
compared to the soft nanopatterned GB junctions. The noise
properties of our nanojunctions allows to identify two classes of
experiments that one can perform by taking advantage of the
specifics of the transport properties:

1) to realize quantum bits by employing soft nanopatterned
junctions. The pristine grain boundary is an ideal candidate to
study the intrinsic source of dissipation in HTS by measuring
relaxation and coherence times in a quantum bit.

2) to realize devices where charging effects are dominant. The large
resistivity values of the conventionally patterned junctions $\rho_N
\simeq 5 \cdot 10^{-7} - 2 \cdot 10^{-6} \Omega$cm$^2$ make these
junctions good candidates for the realization of all-HTS single
electron transistors, which can be used to study possible
subdominant order parameters in HTS materials\cite{Sergey}.


\begin{acknowledgments}
This work has been partially supported by EU STREP project MIDAS,
the Swedish Research Council (VR) under the Linnaeus Center on
Engineered Quantum Systems, the Swedish Research Council (VR) under
the project Fundamental properties of HTS studied by the quantum
dynamics of two level systems and the Knut and Alice Wallenberg
Foundation.
\end{acknowledgments}


\begin{thebibliography}{00}
\bibitem{Koelle1999} D. Koelle, R. Kleiner, F. Ludwig, E. Dantsker, and J. Clarke, Rev. Mod. Phys. \textbf{71}, 631 (1999).
\bibitem{Alff1998} L. Alff, A. Beck, R. Gross, A. Marx, S. Kleefisch,
T. Bauch, H. Sato, M. Naito, and G. Koren, Phys. Rev. B \textbf{58}, 11197 (1998).
\bibitem{Tsuei2000} C. C. Tsuei, and J. R. Kirtley, Rev. Mod. Phys. \textbf{72}, 969 (2000).
\bibitem{Jesper} J. Johansson, K. Cedergren, T. Bauch and F. Lombardi, Phys. Rev. B \textbf{79}, 214513 (2009).
\bibitem{Bauch2005} T. Bauch, F. Lombardi, F. Tafuri, A. Barone, G. Rotoli,
P Delsing, and T. Claeson, Phys. Rev. Lett. \textbf{94}, 087003 (2005).
\bibitem{Bauch2006} T. Bauch, T. Lindstr\"{o}m, F. Tafuri, G. Rotoli,
P. Delsing, T. Claeson, and F. Lombardi, Science \textbf{311}, 57 (2006).
\bibitem{Bauch2009} T. Bauch, D. Gustafsson, K. Cedergren, S. Nawaz, M. Mumtaz Virk,
H. Pettersson, E. Olsson, and F. Lombardi, Phys. Scr. \textbf{T173}, 014006 (2009).
\bibitem{Gustafsson2010} D. Gustafsson, T. Bauch, S. Nawaz, M. Mumtaz Virk,
G. Signorello, and F. Lombardi, Physica C \textbf{470}, S188 (2010).
\bibitem{Kawasaki1992} M. Kawasaki, P. Chaudhari, and A. Gupta, Phys. Rev. Lett. \textbf{68}, 1065 (1992).
\bibitem{Miklich} A.H. Miklich, J. Clarke, M.S. Colclough and K. Char, Appl. Phys. Lett. \textbf{60}, 1899
(1992).
\bibitem{Marx1995a} A. Marx, U. Fath, W. Ludwig, R. Gross and T. Amrein, Phys. Rev. B \textbf{51}, 6735 (1995).
\bibitem{Marx1995b} A. Marx, U. Fath, L. Alff, and R. Gross, Appl. Phys. Lett. \textbf{67}, 1929 (1995).
\bibitem{Marx1998} A. Marx, L. Alff, and R. Gross, Appl. Supercond. \textbf{6}, 621 (1998).
\bibitem{Liatti2006} M. V. Liatti, U. Poppe, and Y. Y. Divin, Appl. Phys. Lett. \textbf{88}, 152504 (2006).
\bibitem{Herbstritt} F. Herbstritt, T. Kemen, L. Alff, A. Marx and R. Gross, Appl. Phys.
Lett. \textbf{78}, 955 (2001).
\bibitem{Davidnanoletter} D. Gustafsson, H. Pettersson, B. Iandolo,
E. Olsson, T. Bauch and F. Lombardi, Nano Lett. \textbf{10}, 4824
(2010).
\bibitem{Komissinski} P.V. Komissinski, B. Högberg, A.Y. Tzalenchuk
and Z. Ivanov, Appl. Phys. Lett. \textbf{80}, 1022 (2002).
\bibitem{tobias} T. Lindström, J. Johansson, T. Bauch, E. Stepantsov, F. Lombardi and S. A. Charlebois, Phys. Rev. B \textbf{74}, 014503
(2006).
\bibitem{DaniellaJAP} D. Stornaiuolo, G. Rotoli, K. Cedergren, D. Born, T. Bauch, F. Lombardi and F.
Tafuri, J. Appl. Phys. \textbf{107}, 113901 (2010).
\bibitem{DaniellaIEEE} D. Stornaiuolo, K. Cedergren, D. Born, T.
Bauch, A. Barone, F. Lombardi and F. Tafuri, IEEE Trans. Appl.
Supercond. \textbf{19}, 174 (2009).
\bibitem{Green} U. Scotti di Uccio, F. Miletto Granozio, A. Di Chiara, F. Tafuri, O.I. Lebedev, K. Verbist and G. van Tendeloo, Physica C \textbf{321},
162 (1999).
\bibitem{DiChiara} A. Di Chiara, et al. Physica C \textbf{273}, 30
(1996)
\bibitem{Tinkham} M. Tinkham, \textit{Introduction to Superconductivity}, 2nd
ed. (Dover Publications, New York, 2004).
\bibitem{Likharev} K. Likharev, \textit{Dynamics of Josephson
junctions and circuits}, 1st ed. (Gordon and Breach Science
publishers, New York, 1986).
\bibitem{Rosenthal} P.A. Rosenthal, M. R. Beasley, K. Char, M.S. Colclough
and G. Zaharchuk, Appl. Phys. Lett. \textbf{59}, 2482 (1991).
\bibitem{VanDuzer} T. Van Duzer and C. W. Turner, \textit{Superconductive devices and
circuits}, 2nd ed. (Prentice Hall, New Jersey, 1999).
\bibitem{Rogers} C.T. Rogers and R.A Buhrman, Phys. Rev. Lett. \textbf{53},
1272 (1984).
\bibitem{Machlup} S. Machlup, J. Appl. Phys. \textbf{24}, 341 (1954).
\bibitem{Gross} R. Gross, L. Alff, A. Beck, O.M. Froehlich, D. Koelle and A. Marx, IEEE Trans. Appl. Supercond. \textbf{7},
2929 (1997).
\bibitem{GrossPRB} R. Gross, P. Chaudhari, M. Kawasaki and A. Gupta, Phys. Rev. B
\textbf{42}, 10735 (1990).
\bibitem{Cedergren} K. Cedergren, H. Pettersson, E. Olsson, T. Bauch and F.
Lombardi, in manuscript.
\bibitem{Lombardi_PhysicaC} F. Lombardi, et al.,Physica C \textbf{435}, 8 (2006)
\bibitem{Mannhart1998} J. Mannhart, and H. Hilgenkamp, Mater. Sci. Eng. B \textbf{56}, 77 (1998).
\bibitem{Hilgenkamp1998} H. Hilgenkamp and J. Mannhart, Appl. Phys. Lett, \textbf{73}, 265 (1998).
\bibitem{Homes} C.C. Homes, S.V. Dordevic, D.A. Bonn, Ruixing Liang, W.N. Hardy and T. Timusk, Phys. Rev. B \textbf{71}, 184515 (2005).
\bibitem{Zuev} Y. Zuev, M.S. Kim and T.R. Lemberger, Phys. Rev. Lett. \textbf{95}, 137002 (2005).
\bibitem{Marx1997} A. Marx, and R. Gross, Appl. Phys. Lett. \textbf{70}, 120 (1997).
\bibitem{VanHarlingen2004} D. J. Van Harlingen, et al., Phys. Rev. B \textbf{70}, 064517 (2004).
\bibitem{Sergey} S. E. Kubatkin, A. Ya. Tzalenchuk, Z. G. Ivanov, P. Delsing, R.I. Shekhter and T. Claeson, JETP Lett. \textbf{63} 126 (1996)

\end{thebibliography}
\end{document}